\documentclass[twocolumn,superscriptaddress,amsmath,amssymb,aps,prb]{revtex4-2}

\usepackage{graphicx}
\usepackage{amsfonts}
\usepackage{amsmath}
\usepackage[T1]{fontenc}
\usepackage{mathtools}
\usepackage{bm}
\usepackage{braket}

\usepackage[dvipsnames]{xcolor}

\usepackage{hyperref}
\hypersetup{colorlinks=true, urlcolor=blue, citecolor = blue}

\begin{document}

\title{First-principles prediction of chiral-phonon-induced orbital accumulation}

\author{A. Pezo}
\email{apezol@issp.u-tokyo.ac.jp}
\affiliation{Institute for Solid State Physics, University of Tokyo, Kashiwa, 277-8581, Japan}

\author{A. Manchon}        
\affiliation{Aix-Marseille Univ, CNRS, CINaM, Marseille, France}

\author{Y. Nii}        
\affiliation{Department of Applied Physics and Physico-Informatics, Keio University, Yokohama 223-8522, Japan}

\author{K. Ando}        
\affiliation{Department of Applied Physics and Physico-Informatics, Keio University, Yokohama 223-8522, Japan}
\affiliation{Keio Institute of Pure and Applied Sciences, Keio University, Yokohama 223-8522, Japan}
\affiliation{Center for Spintronics Research Network, Keio University, Yokohama 223-8522, Japan}

\author{T. Kato}        
\affiliation{Institute for Solid State Physics, University of Tokyo, Kashiwa, 277-8581, Japan}

\date{\today}

\begin{abstract}
Chiral phonons offer a route to transfer angular momentum without relying on magnetic order, but their electronic response in metals remains poorly understood from perspectives beyond spin-based scenarios.
Using first-principles calculations, we show that coherent chiral lattice motion generates orbital accumulation and, through spin-orbit coupling, a smaller accompanying spin accumulation. 
Our approach evaluates orbital and spin expectation values directly from strain-perturbed ab initio Hamiltonians in the long-wavelength limit, where the phonon perturbation is represented by symmetry-adapted circular lattice distortions. 
We show that the response is controlled mainly by orbital character, near-degeneracies, and electron-phonon coupling, rather than by spin-orbit coupling alone. 
These results identify light transition metals as promising platforms for chiral-phonon-driven orbitronics.
\end{abstract}

\maketitle

{\it Introduction.---} The control of spin degrees of freedom in condensed matter lies at the heart of modern nanoscale phenomena and device concepts.
Traditionally, spintronics has focused on spin dynamics and charge--spin conversion mediated by spin--orbit coupling (SOC)~\cite{Manchon2019}. 
More recently, however, collective lattice excitations have emerged as key contributors to angular momentum transport, giving rise to the rapidly developing field of phonon-mediated spintronics~\cite{Zhang2014,Garanin2015,Weiler2011,Weiler2012,Kobayashi2017,Kurimune2020,Funato2024,Qin2025,Yao2025,Nishimura2025}. 
In this framework, lattice vibrations carrying angular momentum, known as \emph{chiral phonons}~\cite{Juraschek2025}, have attracted considerable attention as a means of transferring angular momentum via chiral lattice motion, potentially even in weak-SOC settings~\cite{Ishito2022,Ueda2023,Chaudhary2024}.

Chiral phonons arise in non-centrosymmetric crystals, where degenerate vibrational modes combine into circularly polarized lattice motions with well-defined angular momentum and an associated pseudospin degree of freedom~\cite{Juraschek2025,Ueda2023,Ishito2022,Chaudhary2024,Ohe2024}. Unlike conventional phonons, their circular atomic motion can couple selectively to electronic, magnetic, and valley degrees of freedom~\cite{Maity2022,Basini2024}. This versatility has motivated their exploration across a wide range of systems, including ferroelectrics~\cite{Chen2024,Gao2023}, topological materials~\cite{Hernandez2023}, and magnetic systems where coupling to magnons can generate hybrid excitations~\cite{Weissenhofer2025}. Through SOC, phonon angular momentum can be transferred to electronic spins, leading to effects such as spin polarization, spin relaxation, and spin current generation~\cite{Fransson2023}, thus opening pathways toward low-dissipation spintronic devices driven by lattice dynamics.

While most of these developments emphasize spin-mediated responses, the inherent presence of orbital angular momentum in solids suggests an alternative route for angular momentum transfer that does not rely on strong SOC. Recent theoretical and experimental studies have established orbital transport as a central ingredient of nonequilibrium angular-momentum dynamics in solids~\cite{Jo2024,Cysne2025,Pezo2022,Han2025,Choi2023,Ando2025Review}. In particular, recent experiments reported chiral-phonon generation of orbital currents in light transition metals~\cite{Rovirola2025,Wu2025,Taniguchi2025}, providing a direct experimental evidence for the orbital response investigated in this work.

\begin{figure}[tb]
    \centering
    \includegraphics[width=0.9\linewidth]{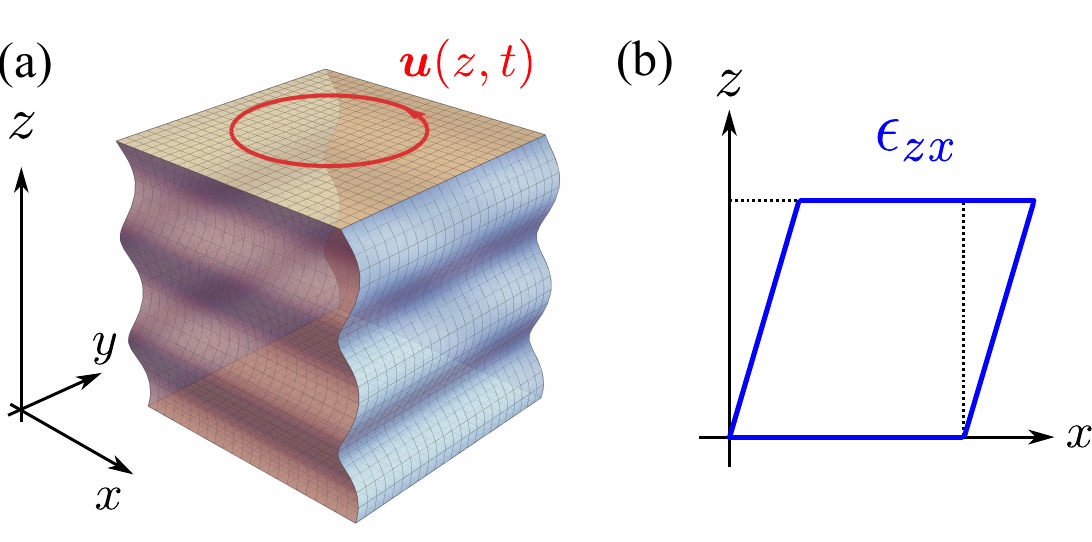}
    \caption{Schematic illustration of the chiral lattice excitation considered in this work. (a) Chiral long-wavelength phonon propagating in a nonmagnetic target (NM) along the $z$ direction. The lattice displacement is represented by circular motion. (b) Schematic of the induced strain $\epsilon_{zx}$ due to the chiral phonon.}  
    \label{Fig:pictorial}
\end{figure}

These developments naturally raise a materials-oriented question: which metals develop orbital accumulation under chiral lattice motion, and what microscopic ingredients control its magnitude? To address this problem, we evaluate orbital and spin expectation values directly from self-consistent first-principles Hamiltonians under symmetry-adapted lattice distortions that represent the long-wavelength, adiabatic limit of coherent chiral phonons (see Fig.~\ref{Fig:pictorial}(a)). In this construction, the circular motion is not treated as a spontaneous bulk mode of the centrosymmetric metals themselves, but as an externally imposed perturbation transmitted from a chiral phonon source to the target material.

This strategy bridges the intuitive picture introduced in recent theoretical studies of phonon-induced orbital accumulation~\cite{Rovirola2025,Sato2025,Yao2025b} with a realistic electronic-structure treatment. 
It is also closely connected to recent experimental reports of chiral-phonon-induced orbital currents in light transition metals~\cite{Rovirola2025,Wu2025,Taniguchi2025}.
We show that the induced orbital accumulation is generally much larger than the accompanying spin accumulation, and that its magnitude is governed not by SOC alone but by the combination of orbital texture, near-degeneracies, and electron-phonon coupling. As a consequence, several weak-SOC transition metals show a comparable or even stronger orbital response to the same chiral lattice perturbation than heavier metals commonly used in spintronics.

{\it Formulation.---} We consider the long-wavelength limit of a coherent transverse acoustic chiral phonon injected into the target material. In this limit, the ionic motion is described by a smooth displacement field $\bm{u}(\bm{r},t)$, and the relevant lattice perturbation acting on the electrons is governed by its gradient. The symmetric part of the displacement gradient defines the strain tensor,
\begin{equation}
\epsilon_{ij}(\bm{r},t)=\frac{1}{2}\left(\partial_i u_j+\partial_j u_i\right).
\label{eq:strain_tensor_main}
\end{equation}
For a sufficiently long-wavelength acoustic mode, the electronic response is therefore fully captured by a strain-dependent Hamiltonian.

Within the adiabatic approximation, the strain perturbation is recast as a dependence of the first-principles Hamiltonian on the homogeneous strain components,
\begin{equation}
H[\epsilon] = H_0 + \sum_{ij}\epsilon_{ij}\,D_{ij} +\mathcal{O}(\epsilon^2),
\label{eq:strain_hamiltonian_main}
\end{equation}
where $D_{ij} = \partial H/\partial \epsilon_{ij}|_{\epsilon=0}$ is the deformation operator associated with the strain channel $\epsilon_{ij}$. 
Physically, $D_{ij}$ describes how strain modifies onsite crystal fields, orbital hybridization, and intersite hopping. In the present framework, the primary response is orbital, while spin accumulation arises from the conversion of orbital angular momentum through intra-atomic SOC.

These deformation operators are evaluated from finite differences between strained self-consistent \textit{ab initio} Hamiltonians,
\begin{equation}
D_{ij}
\approx
\frac{H(\epsilon_{ij})-H(-\epsilon_{ij})}{2\epsilon_{ij}}. 
\label{eq:finite_difference_main}
\end{equation}
Since a uniform strain corresponds to a displacement field linear in position, $u_j(\bm{r})=\epsilon_{ji}r_i$, the actual atomic displacements are proportional to the lattice vectors and internal coordinates. In the nonorthogonal SIESTA implementation, the effective matrix elements entering the Berry-curvature formulas are constructed from $\partial_i H$ and $\partial_i S$, where $S$ is the overlap matrix for the nonorthogonal basis set.
For details, see the Supplemental Material~\cite{Supplement}.

To model a chiral acoustic perturbation, we consider two orthogonal transverse lattice distortions in the plane perpendicular to the propagation direction and combine them with a phase shift of $\pi/2$, as shown schematically in Fig.~\ref{Fig:pictorial}(a). For cubic materials, we take the propagation direction along the $[001]$ axis, while for hcp materials we take it along the $c$ axis. 
In this geometry, the displacement field takes the form
\begin{equation}
\bm{u}(\bm{r},t)= \left(\delta \cos (\omega t - q_z z),\lambda \delta \sin (\omega t - q_z z),0\right),
\end{equation}
where $\delta$ is the amplitude of the lattice displacement and $\lambda$ denotes the chirality of the transverse phonon mode. Locally, this traveling-wave modulation defines an adiabatic strain cycle, so that the long-wavelength phonon can be treated within the same strain-dependent Hamiltonian. The natural adiabatic driving parameters are then the two shear strain components
\begin{align}
& \epsilon_{zx}(z,t)=\frac{1}{2}\partial_z u_x = \frac{\delta q_z}{2} \sin (\omega t - q_z z), \label{eq:strain_channels_main1}\\
& \epsilon_{zy}(z,t)= \frac{1}{2}\partial_z u_y = - \frac{\lambda \delta q_z}{2} \cos (\omega t - q_z z).
\label{eq:strain_channels_main2}
\end{align}
We show a schematic image for the strain $\epsilon_{zx}$ in Fig.~\ref{Fig:pictorial}(b). We focus on the spin and orbital accumulation around the fixed position $z = z_0$, for which the Hamiltonian with the adiabatic cycle is written as
\begin{equation}
H(t)\approx H_0+\epsilon_{zx}(z_0,t)D_{zx}+\epsilon_{zy}(z_0,t)D_{zy} .
\label{eq:driven_hamiltonian_main}
\end{equation}
This construction should be viewed as a common long-wavelength driving protocol for material comparison, rather than as an explicit calculation of the full phonon eigenproblem of each crystal.

The central observables in the present work are the orbital and spin expectation values evaluated directly from the instantaneous strained electronic states,
\begin{align}
\langle \hat{L}_z \rangle(t)
&= \sum_{n\bm{k}} f_{n\bm{k}}
\langle \psi_{n\bm{k}}(t)|\hat{L}_z|\psi_{n\bm{k}}(t)\rangle,
\label{eq:Lz_direct_main} \\
\langle \hat{S}_z \rangle(t)
&= \sum_{n\bm{k}} f_{n\bm{k}}
\langle \psi_{n\bm{k}}(t)|\hat{S}_z|\psi_{n\bm{k}}(t)\rangle,
\label{eq:Sz_direct_main}
\end{align}
where $|\psi_{n\bm{k}}(t)\rangle$ are the eigenstates of the strain-dependent Hamiltonian in Eq.~(\ref{eq:driven_hamiltonian_main}). 
In this formulation, the orbital accumulation is obtained directly from the strained first-principles electronic structure, while the spin accumulation follows from spin-orbit-mediated conversion of the orbital response.

The cycle-averaged chiral contribution is a directed quantity and, to leading order in $\omega$, is bilinear in the two orthogonal strain amplitudes,
\begin{equation}
\overline{\langle \hat{O} \rangle}_{\mathrm{chiral}}
=
\chi_{zx,zy}^{(O)}
\left(
\dot{\epsilon}_{zx} \epsilon_{zy}-\dot{\epsilon}_{zy} \epsilon_{zx}
\right),
\label{eq:chiral_response_main}
\end{equation}
with $\hat{O}=\hat{L}_z$ or $\hat{S}_z$. For the circular drive in Eq.~\eqref{eq:strain_channels_main1} and \eqref{eq:strain_channels_main2}, this becomes
\begin{equation}
\overline{\langle \hat{O} \rangle}_{\mathrm{chiral}}
= -\frac{\lambda \delta^2 q_z^2\omega}{4}  \chi_{zx,zy}^{(O)},
\label{eq:quadratic_scaling_main}
\end{equation}
showing explicitly that the leading chiral dc response is quadratic in the deformation amplitude. Here, $\chi_{zx,zy}^{(O)}$ is the intrinsic strain-response coefficient extracted from the cycle-averaged direct expectation values. It also admits a compact geometric interpretation in the adiabatic strain-parameter space,
\begin{equation}
\chi_{zx,zy}^{(O)}
=
\int_{\mathrm{BZ}}\frac{d\bm{k}}{(2\pi)^d}\sum_n^{\rm occ}
\Omega_{zx,zy}^{(O,n)}(\bm{k}),
\label{eq:chi_berry_main}
\end{equation}
where $\Omega_{zx,zy}^{(O,n)}(\bm{k})$ is the generalized Berry curvature associated with the observable $\hat O$ for band $n$ in the strain space spanned by $\epsilon_{zx}$ and $\epsilon_{zy}$. Thus, the chiral response is controlled by both the oriented area enclosed by the adiabatic cycle and the geometric structure of the occupied electronic states. The detailed derivation of Eq.~(\ref{eq:chi_berry_main}) and its implementation in SIESTA are given in Supplemental Material~\cite{Supplement} and in related model analyses~\cite{Yao2025b,Sato2025}.

\begin{figure}
    \centering
    \includegraphics[width=\linewidth]{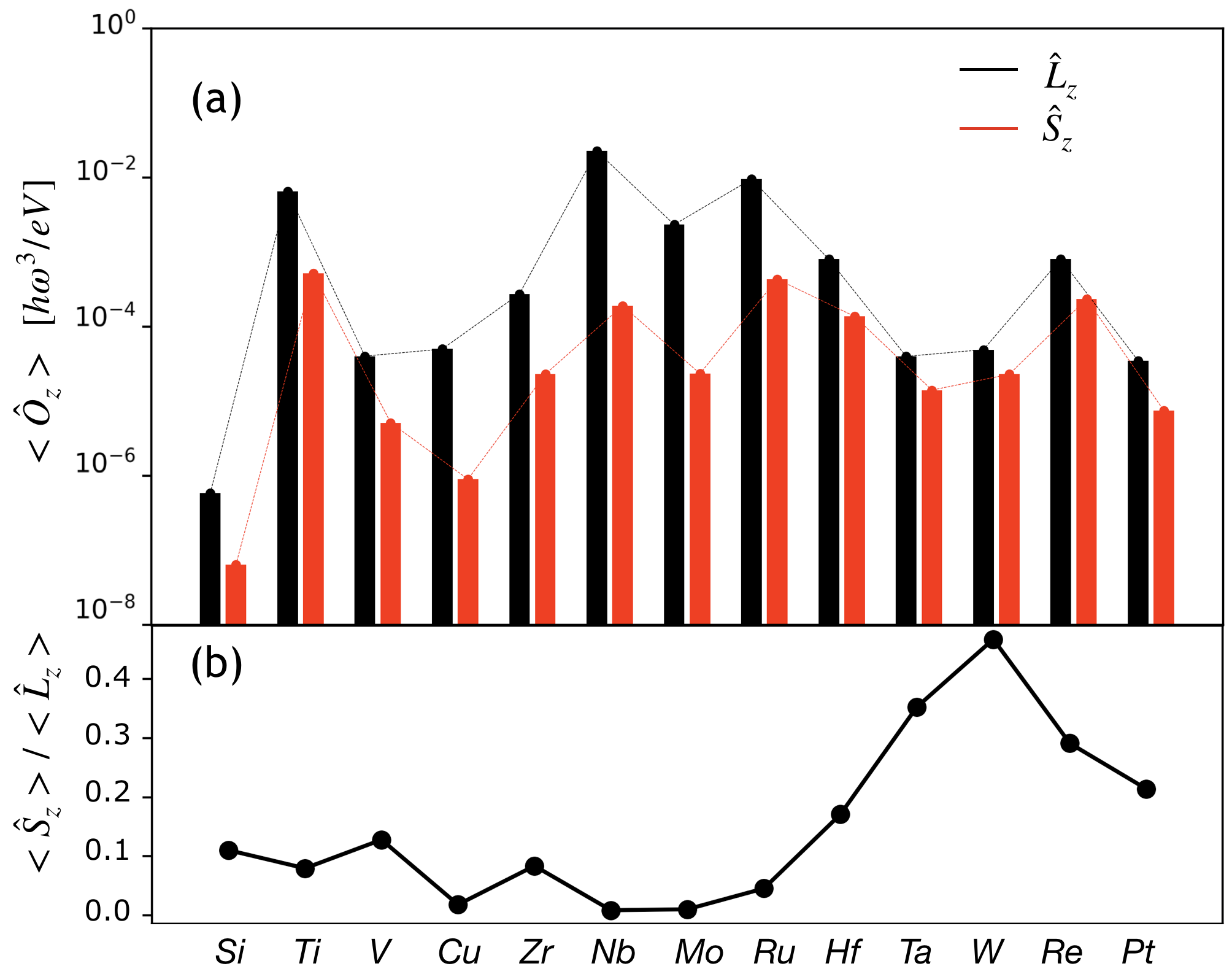} 
    \caption{Orbital and spin accumulations induced by the adiabatic chiral strain cycle. (a) Orbital and spin accumulations for the materials considered in this work. Several light transition metals exhibit larger orbital accumulation than Pt despite its stronger SOC. (b) Ratio $\braket{\hat{S}_z}/\braket{\hat{L}_z}$, illustrating more efficient orbital-to-spin conversion in heavier metals.}
\label{Fig:accumulation_metals}
\end{figure}

{\it Results.---} We apply the above formulation to a set of representative transition metals, together with Si as a reference system, and evaluate the orbital and spin accumulations generated by the fixed amplitude of the lattice displacement, $\delta = 10^{-3}\,$\AA.
The resulting orbital and spin accumulations are shown in Fig.~\ref{Fig:accumulation_metals}(a).
The first general result is that the orbital accumulation is systematically larger than the accompanying spin accumulation. 
This feature originates from the fact that the primary effect of the chiral lattice deformation is orbital, because the strain directly modulates orbital hybridization and hopping amplitudes, whereas the spin signal emerges only through SOC-mediated conversion.
The second key result is that the orbital response varies strongly across materials. Among the systems considered here, Ti, Nb, Mo, and Ru display the largest orbital accumulations, whereas the response for Si, Cu, and Pt remains comparatively small.

To assess the role of SOC more directly, Fig.~\ref{Fig:accumulation_metals}(b) shows the ratio $\langle \hat{S}_z\rangle/\langle \hat{L}_z\rangle$ for the different materials. As expected, this ratio tends to be larger for heavy elements such as Ta and W, indicating more efficient orbital-to-spin conversion.
At the same time, this panel clarifies that strong SOC alone is not sufficient to maximize the total response: Pt, despite its large SOC, exhibits only a weak orbital accumulation and therefore a comparatively modest overall spin signal. 
We also point out that the ratio $\langle \hat{S}_z\rangle/\langle \hat{L}_z\rangle$ varies moderately across materials, ranging from $\sim 0.01$ for Nb to $\sim 0.5$ for W.
These results suggest that the dominant material descriptor is not SOC by itself, but the interplay between strain-induced electronic mixing, orbital character, and near-degeneracy of the relevant states.
Indeed, from the explicit expression of the Berry curvature~\cite{Supplement}, it is expected to be strongly enhanced when orbital degeneracy lies near the Fermi energy~\cite{Sato2025}.

\begin{figure}
\centering
\includegraphics[width=\linewidth]{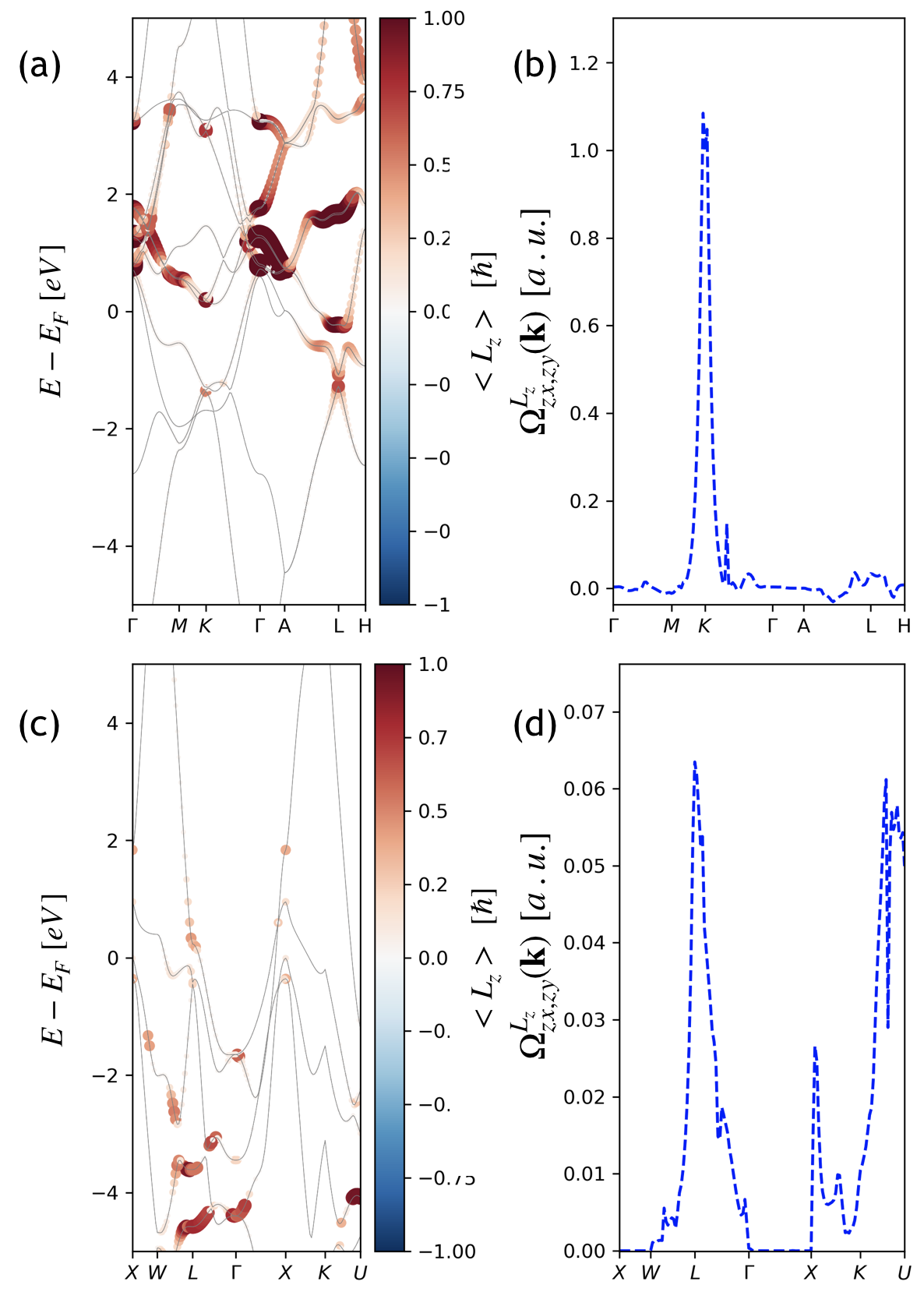}
\caption{Microscopic origin of the contrasting responses in Ti and Pt. Orbital texture projected onto the band structures of Ti (a) and Pt (c), together with the corresponding electron-phonon and orbital Berry-curvature indicators in Ti (b) and Pt (d). Ti exhibits stronger orbital texture, more favorable near-degeneracies, and a larger strain-induced response than Pt despite its weaker SOC.}   \label{Fig:energy_displacement}
\end{figure}

To examine the microscopic origin of the response in more detail, Fig.~\ref{Fig:energy_displacement}(a,c) compares the band structures of Ti and Pt projected onto the orbital angular momentum. Ti exhibits relatively localized $d$-orbital character, weaker band dispersion around the Fermi level, and stronger orbital texture near nearly degenerate regions. These features favor strain-induced inter-orbital mixing and are accompanied by pronounced peaks in the generalized Berry curvature associated with the orbital accumulation [Fig.~\ref{Fig:energy_displacement}(b)]. By contrast, Pt represents the opposite limit: although SOC is strong, the relevant bands are more dispersive, the orbital texture is less favorable, and the corresponding response generated by the same chiral deformation is much smaller [Fig.~\ref{Fig:energy_displacement}(d)]. The Ti/Pt comparison therefore highlights the hierarchy of ingredients controlling the response: orbital texture and near-degeneracy set the strength of strain-induced electronic mixing, whereas SOC mainly governs the subsequent conversion of orbital angular momentum into spin.

{\it Discussion.---} The comparison across materials supports a simple physical picture. Efficient chiral-phonon-driven accumulation requires, first, orbitally active bands that respond strongly to strain, and second, sufficiently strong electron-phonon coupling to convert the imposed circular lattice motion into orbital polarization. Strong SOC is not the main requirement for generating the response itself, but it does determine how much of that orbital accumulation is subsequently transferred into spin. This is why weak-SOC transition metals such as Ti or Nb are better orbital generators than Pt, while heavier metals such as Ta and W remain more efficient at orbital-to-spin conversion once the orbital accumulation is established.

From an experimental perspective, the predicted orbital accumulations are of the order of $10^{-5}\mu_{\mathrm{B}}$ per atom ($\mu_{\mathrm{B}}$: the Bohr magneton) for the most favorable materials considered here.
The voltage signal due to the inverse orbital Hall effect or the inverse orbital Rashba-Edelstein effect is roughly estimated by $\langle \hat{L}_z \rangle/e\mathcal{N}(E_F)$, where $\mathcal{N}(E_F)$ is the density of states at the Fermi energy.
Using our numerical results, the voltage signal is estimated to be of order $\mu\mathrm{eV}$, which should already be relevant for present orbitronic detection schemes. The same framework can be extended to systems where symmetry, magnetism, or reduced dimensionality further enhance orbital texture, including ferromagnets, ferroelectrics, and low-dimensional materials.

\begin{figure}
\centering
\includegraphics[width=1.0\linewidth]{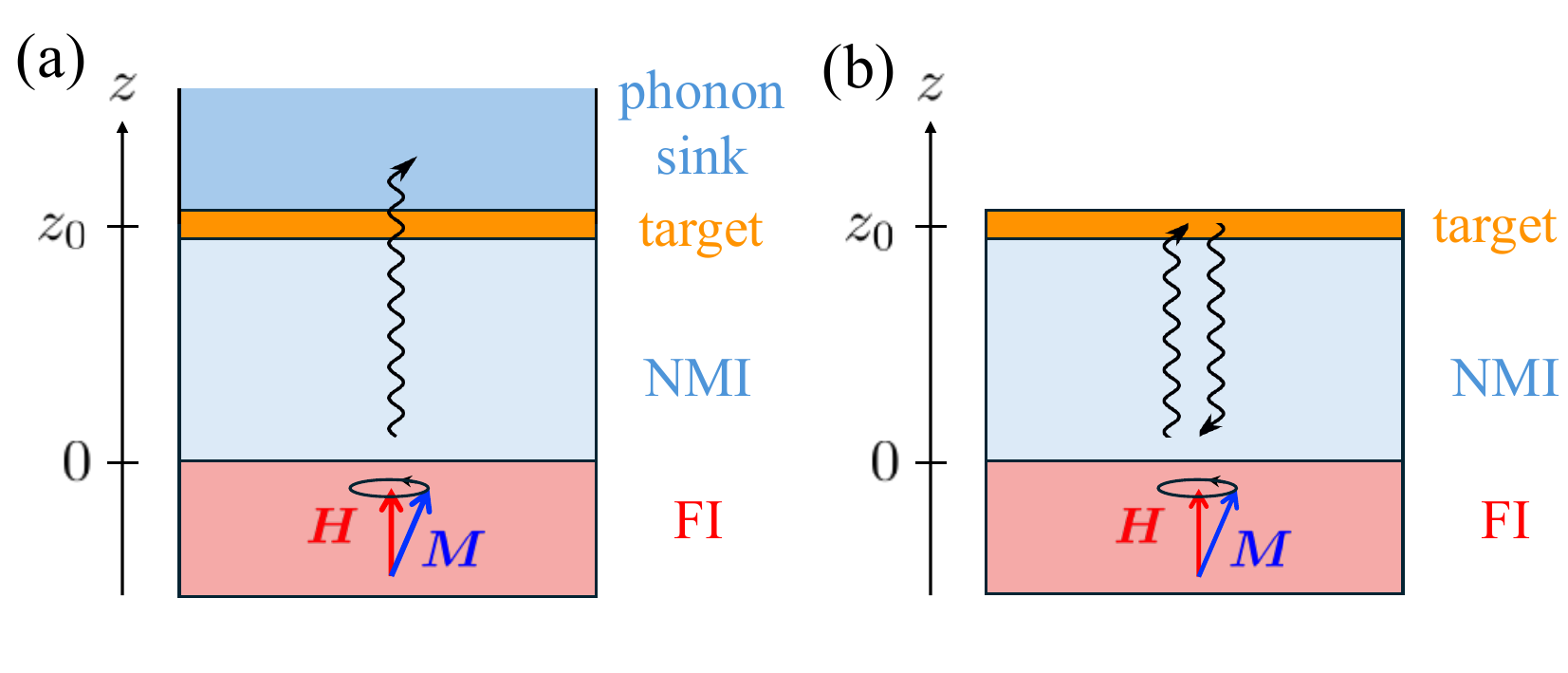}
\caption{Schematic experimental setups for detecting orbital and spin accumulation induced by chiral acoustic phonon, using a ferromagnetic insulator (FI) and a nonmagnetic insulator (NMI). (a) Injection of a propagating chiral phonon mode into a target. (b) Standing wave of a chiral phonon with the target placed on the nonmagnetic insulating spacer.}   \label{Fig:Exp}
\end{figure}

{\it Experimental realization.---}
One possible route to realizing the required chiral lattice driving is to use magnetoelastic generation of transverse elastic waves in a magnetic insulator~\cite{Bommel1959,Comstock1963,Seavey1965,Kittel1958,GurevichMelkov1996}, an approach recently revisited in the context of spintronic functionality~\cite{Streib2018,An2020}. A natural heterostructure platform consists of a magnetic insulator such as YIG, a nonmagnetic insulating spacer such as GGG, and a target metal [Fig.~\ref{Fig:Exp}(a)]. When the magnetization of the magnetic layer is driven into precession by microwaves, the magnetoelastic coupling launches a chiral transverse phonon toward the target stack. If the target metal is thinner than the acoustic wavelength, the lattice displacement, and hence the induced orbital accumulation, are approximately uniform across the metal thickness.

An alternative implementation is provided by a standing-wave geometry in a finite-thickness structure [Fig.~\ref{Fig:Exp}(b)]. In this case, the local strain amplitude depends on the position of the target metal within the standing-wave profile, and the response is maximized when the metal is placed near an antinode of the relevant shear strain. The detailed node--antinode structure depends on the acoustic boundary conditions, but the same local-response picture applies once a standing shear mode is established.

In either geometry, the resulting orbital accumulation could be detected electrically by the inverse orbital Rashba--Edelstein effect~\cite{ElHamdi2023,Hayashi2024} or optically by MOKE~\cite{Choi2023,Lyalin2023}. In the nearly uniform thin-film regime, inverse orbital Rashba--Edelstein detection appears particularly natural. In contrast, inverse orbital Hall detection in the thin-film limit would more likely rely on lateral diffusion of the accumulated orbital angular momentum in an extended metal layer, followed by charge conversion in a suitable orbital-charge converter.

{\it Summary.---} In summary, we have developed a first-principles framework to calculate orbital and spin accumulations induced by coherent chiral lattice motion in metals. The main outcome is that orbital accumulation is the dominant response and is maximized not in the strongest-SOC materials, but in systems where orbital texture, near-degeneracy, and electron-phonon coupling cooperate most efficiently. This explains why several light transition metals are stronger candidates than Pt for chiral-phonon-driven orbital generation, while heavier metals remain more efficient at orbital-to-spin conversion. Our results therefore provide a concrete materials guide for orbitronic experiments driven by nonequilibrium lattice angular momentum.
 
\begin{acknowledgments}
The authors thank T. Sato and T. Sohier for helpful discussions. 
This work was supported by Grants-in-Aid for Scientific Research (Grants No. JP23KJ0702 and No. JP24K06951) and the Japan Science and Technology Agency (JST) ASPIRE Program No. JPMJAP2410.
A.M. was supported by the EIC Pathfinder OPEN grant
101129641 ``OBELIX'', and by France 2030 government
investment plan managed by the French National Research Agency under grant reference PEPR SPIN--[SPINTHEORY] ANR-22-EXSP-0009 and [OXIMOR] ANR-24-EXSP-0011.
K.A. acknowledges support from JSPS KAKENHI (Grant Nos. JP22H04964 and JP25K21707), the Spintronics Research Network of Japan (Spin-RNJ), and the MEXT Initiative to Establish Next-generation Novel Integrated Circuits Centers (X-NICS) (Grant No. JPJ011438).
Y.N. acknoledges support from JSPS KAKENHI (Grant No. JP24K21725 and JP24K01283), and JST-FOREST (Grant No. 	JPMJFR2133).
\end{acknowledgments}

\bibliography{ref}

\end{document}